\documentstyle[prl,aps,preprint,axodraw,12pt]{revtex}
\tightenlines
\input{epsf.sty}
\input{psfig.sty}

\newcommand{\beqq}{\begin{equation}}
\newcommand{\eeqq}{\end{equation}}
\newcommand{\beq}{\begin{eqnarray}}
\newcommand{\beas}{\begin{eqnarray*}}
\newcommand{\eeq}{\end{eqnarray}}
\newcommand{\eeas}{\end{eqnarray*}}
\newcommand{\ba}{\begin{array}}
\newcommand{\ea}{\end{array}}
\begin{document}

\draft 
\preprint{\vbox{ \hbox{IFT-P.022/2001}}}

\title{Scalar scenarios contributing to $(g-2)_{\mu}$ with enhanced Yukawa
couplings}
\author{C. A. de S. Pires\footnote{e-mail:cpires@ift.unesp.br} and
P. S. Rodrigues da Silva\footnote{e-mail:fedel@ift.unesp.br}}
\address{Instituto de F\'\i sica Te\'orica,
Universidade Estadual Paulista.\\
Rua Pamplona, 145
01405-900-- S\~ao Paulo, SP
Brazil.}
\date{March, 2001}

\maketitle
\begin{abstract}
{ In this work we address contributions from scalars to
$(g-2)_\mu$. In order to explain the recently measured deviation
by the BNL experiment on $(g-2)_\mu$, it is necessary that these
scalars are either light or couple strongly with muons. Here we
explore this last possibility. We show that a scalar with mass of
the order of $10^2$~GeV provides significant contribution to
$(g-2)_\mu$ if the Yukawa coupling is about $10^{-1}$. We suggest
scenarios where this comes about naturally.}

\end{abstract}
\newpage

%%%%%%%%%%%%%%%%%%%%%%%%%%%%%%%%%%%%%%%%%%%%%%%%%%%%%%%%%%%%%%%%%%%%%%
\section{introduction}

A new measurement of the muon anomalous magnetic moment,
$(g-2)_\mu$, was recently announced which indicates a deviation
from the theoretical value of 2.6 sigma\cite{brown}
\beqq a^{exp}_\mu - a^{SM}_\mu = 426\pm 165 \times 10^{-11}.
\label{deviation} \eeqq
If this result persists~\cite{Yndu} it implies an exciting window
requiring new physics beyond Standard Model (SM). In order to
explain such a deviation, various scenarios have been proposed:
supersymmetry, new gauge bosons, leptoquarks,
etc~\cite{variosscenarios,ma1}. So far, SM extensions in the
scalar sector have been almost neglected, which is at least
plausible since in the SM the interactions between the muon and
the neutral Higgs, $H$,
\beqq f_{\mu \mu} \bar{\mu} \mu H, \label{yakawa1} \eeqq
gives the following contribution to $(g-2)_\mu$~\cite{moore},
\beqq a^\mu_H =\frac{f_{\mu \mu}^2 m_\mu^2}{12 \pi^2 m^2_H}.
\label{hcont} \eeqq
Here, $f_{\mu\mu}$ is the usual Yukawa coupling for the muon, and
has the following form
\beqq f_{\mu \mu} = \frac{m_\mu}{v_w} , \label{fmumu1} \eeqq
where, $m_\mu = 0.105$~GeV, is the muon mass and $v_w = 247$~GeV
is the vacuum expectation value (vev) of the scalar doublet in
the SM. These lead to,
\beqq f_{\mu \mu} \sim 10^{-3}. \label{fmumu2} \eeqq
With this value for $ f_{\mu \mu}$ and considering the Higgs mass
of the order of hundreds of GeV, $m_H \sim 10^2$~GeV, the standard
Higgs contribution to $(g-2)_{\mu}$ is negligible,
\beqq a^\mu_H \sim 10^{-13}. \label{smhc} \eeqq

The previous analysis is general enough to be extended to other
scalar scenarios. However, if we want to generate $a^\mu \sim
10^{-9}$ through scalars, we have to consider either a light
scalar of mass around $m_H \sim 10$~GeV and $f_{\mu \mu} \sim
10^{-2}$, or keep $m_H \sim 10^2$~GeV and take $f_{\mu \mu} \sim
10^{-1}$, which demands a vev, $v\sim 1$~GeV. This last
possibility is very suggestive, since a VEV of the order of few
GeV's is a natural choice when this scalar is in charge of
generating the charged lepton masses, and the heaviest lepton,
the tau, has mass $m_\tau \sim 1.7$~GeV. In view of this we
expect that economic modifications on the scalar sector in SM can
lead to such a favorable configuration of parameters.

Motivated by this we want to suggest scenarios where this can
come about, like minimal extensions in the scalar sector of SM
itself (adding more scalars), as well as a model that requires
small vev's, like $3-3-1$~\cite{pires}.

In what follows, we will present the scenarios we have in mind. In
section \ref{sec1} we extend the SM scalar sector by adding a
doubly charged Higgs boson, which interacts solely with right
handed charged leptons. We compute its contribution to
$(g-2)_\mu$ for a range of masses fixing the Yukawa coupling, so
that we can choose the appropriate mass  for the pointed
deviation in $(g-2)_\mu$. In section \ref{sec2} we introduce a
second Higgs doublet interacting with leptons only, which
generates the charged lepton masses. We present the potential for
the scalars and, considering the constraints coming from it, as
well some suitable choices of the remaining parameters, we are
able to find  the mass which best fits the deviation. We then
embed, in section \ref{sec3}, the previous scenarios in a $3-3-1$
model. Finally, we present some concluding remarks in section
\ref{sec4}.

\section{Doubly charged scalar}
\label{sec1}

We first consider a minimal extension of the scalar sector in the
SM in order to accommodate a doubly charged scalar singlet
$\eta^{++}$. We also attribute to it two units of lepton number
$L=L_e + L_\mu + L_\tau =-2$ and hypercharge $Y=4$, so that it
interacts only with the right-handed leptons as follows,
\beqq {\cal L} = h_{ab} l_{aR}Cl_{bR} \eta^{++} \label{eta}, \eeqq
where $C$ is the charge conjugation matrix in some representation.
This interaction provides six additional contributions to
$(g-2)_\mu$ besides those from SM.

However in what follows we just take into account  the
contributions that conserve flavor, i.e., those which involve the
coupling $h_{\mu \mu}$. This seems a natural assumption since
these contributions are expected to dominate over the
off-diagonal ones. Let us postpone the discussion about the lepton
flavor mixing terms and compute the $(g-2)_\mu$ for the diagonal
term only, assuming a strongly coupled $\eta^{++}$ to leptons.
The two contributions considered here are depicted in Figure (1),
and can be expressed respectively by,
\beq &&a)\;\;\;\;a^{\eta}_\mu = \frac{-h^2}{2\pi^2} \int_0^1
\frac{x^3 -x^2}{x^2 + (z-2)x +1} ,
\nonumber \\
&&b)\;\;\;\; a^{\eta}_\mu = \frac{h^2}{4\pi^2} \int_0^1 \frac{x^2
-x^3}{x^2 + z(1-x)}, \label{eta++cont} \eeq
where $z=\frac{m^2_\eta}{m^2_\mu}$ and $h=h_{\mu \mu}$.

Considering $h \simeq 1$, we observe that the measured deviation
$a_\mu \sim 10^{-9}$ favors such a doubly charged scalar with a
mass $m_\eta \sim 200$~GeV. This is very interesting, since the
addition of a singlet scalar is the simplest modification we can
imagine in the scalar sector of the SM. Of course, one could
imagine such a scalar as an ingredient in some classes of models
dealing with more fundamental questions besides the muon
anomalous magnetic moment. It is reasonable, then, to expect that
this economic addition to SM could be embedded in a larger
structure, and it is in this context that we hope this extension
plays an important role, although here we only worried in
suggesting a picture where heavy scalars would be important for
the $(g-2)_\mu$ deviation.
\begin{figure}
\centerline{ \epsfxsize=0.30\textwidth
\begin{picture}(100,240)(-45,50)
\put(-78,165){\makebox(0,0)[br]{a)}}
\put(-58,165){\makebox(0,0)[br]{$\mu^-$}}
\put(27,165){\makebox(80,0)[br]{$\mu^-$}}
\put(10,195){\makebox(80,0)[br]{$\eta^{++}$}}
\put(20,235){\makebox(20,235)[br]{$\gamma$}}
\ArrowLine(-55,165)(85,165) \DashArrowArc(15,165)(50,0,180)5
\Photon(15,215)(65,240)3 5
\put(-78,85){\makebox(0,0)[br]{b)}}
\put(-58,85){\makebox(0,0)[br]{$\mu^-$}}
\put(27,85){\makebox(80,0)[br]{$\mu^-$}}
\put(10,115){\makebox(80,0)[br]{$\eta^{++}$}}
\put(20,60){\makebox(20,60)[br]{$\gamma$}}
\ArrowLine(-55,85)(85,85) \DashArrowArc(15,85)(50,0,180)5
\Photon(15,85)(65,60)3 5
\end{picture}}
\caption{Doubly charged scalar contributions to the muon
anomalous magnetic moment $(g-2)_\mu$.} \label{fig1}
\end{figure}
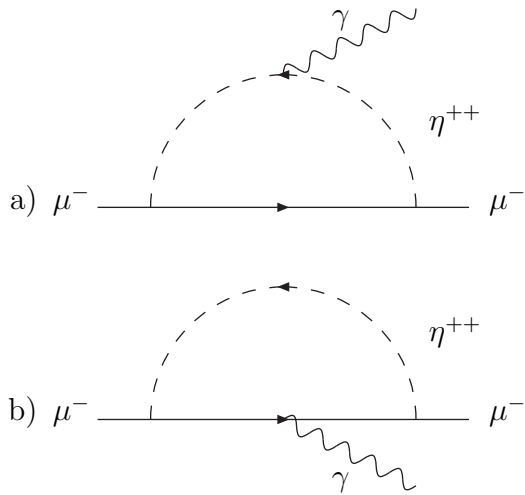

We can turn now to the point concerning the assumption we made
over the off-diagonal couplings in Eq.(\ref{eta}). Those
interactions clearly lead to flavor changing  processes by
exchanging $\eta^{++}$. Such processes severely constrain the
off-diagonal components of the Yukawa coupling matrix, $h_{ab}$
for $a\neq b$. Let us consider only three of the flavor changing
process\footnote{Other rare leptonic decays, for each lepton
flavor decaying in three leptons, are of the same order of
magnitude as those presented here and would lead to the same
constraints.} : $\mu \rightarrow 3e$, $\tau\rightarrow
3\mu\,,\,\,3e$. The decay rate of a lepton, $l^{\prime}$, in
three lighter leptons, $l$, allowed by the interaction in
Eq.~(\ref{eta}) has, in general, the following
expression~\cite{babu}

\beq \Gamma(l^{\prime} \rightarrow 3l)\simeq
\frac{h^2_{l^{\prime} l}
h^2_{ll}}{192\pi^3}\frac{m^5_{l^{\prime}}}{m^4_\eta}.
\label{rate}
\eeq
The present experimental data on these flavor changing processes
are: $BR(\mu \rightarrow 3e) \lesssim 10^{-12}$, $BR(\tau
\rightarrow 3e, 3\mu) \lesssim 10^{-6}$\cite{pdg}. This can be
translated to the following constraints: $\frac{h_{e \mu}
h_{ee}}{m^2_\eta}\lesssim 10^{-11} $~GeV$^{-2}$ and $\frac{h_{e
\tau} h_{ee}}{m^2_\eta}, \frac{h_{\mu \tau} h_{\mu \mu}}{m^2_\eta}
\lesssim 10^{-7}$~GeV$^{-2}$. If we have a scalar with mass,
$m_\eta \simeq 10^2$~GeV, these constraints require:
$h_{e\mu}h_{ee}\lesssim 10^{-7}$ and
$h_{e\tau}h_{ee},\,\,h_{\mu\tau}h_{\mu\mu} \lesssim 10^{-3}$.
Hence in order to explain the $(g-2)_\mu$ with heavy scalars, and
consequently, enhanced diagonal Yukawa couplings, $h_{aa}$, we
can safely assume $h_{a\neq b}\simeq 0$.  By the other hand, if
we demand that the interaction in Eq.(\ref{eta}) respects all the
global symmetries of the standard model, we automatically have
$h_{a\neq b} =0$ since the off-diagonal components of $h$ violate
the global symmetries $U(1)_{L_e,L_\mu ,L_\tau}$.

Concerning the diagonal components, there is a lower bound to the
product of $h_{ee}$ and $h_{\mu\mu}$ imposed by
muonion-antimuoniun conversion: $\frac{h_{ee}h_{\mu
\mu}}{m^2_\eta}>10^{-8}$~GeV$^{-2}$~\cite{mohabook}. For $m_\eta
\simeq 10^2$~GeV we have $h_{ee}h_{\mu \mu}>10^{-4}$. Along with
this the only upper bounds come from $(g-2)_e$ and the Bhabha
scattering process. From these,  the last is the most
stringent~\cite{bhabha}: $\frac{h_{ee}^2}{m^2_\eta}
<10^{-6}$~GeV$^{-2}$, which requires, in our case, $h_{ee}<
10^{-1}$. There is no experimental constraint on $h_{\mu \mu}$,
except for the recent $(g-2)_\mu$ deviation, which can be solved,
as in the above proposal, with $h_{\mu \mu} \simeq 1$ for $m_\eta
\simeq 10^{2}$~GeV.

\section{a second scalar doublet}
\label{sec2}

The second scenario we consider here is the one where a
second Higgs doublet,
\beq H_1 = \left (
\begin{array}{c}
H_1^+ \\
H_1^0
\end{array}
\right )
\label{secdoub},
\eeq
is added to the SM. Let us assume that this Higgs is in charge of
generating charged lepton masses only,
\beq {\cal L}^Y = f_{ii} \bar L_{iL} H_1 l_{iR} + h.c.
\label{yakgen} \eeq

Now, consider the following potential for the two Higgs doublet:
\beq V=\frac{\mu^2_1}{2} H^2_1 + \frac{\mu^2_2}{2}H^2_2 +
\frac{\lambda_1}{4} H^4_1 + \frac{\lambda_2}{4} H^4_2 +
\frac{\lambda_3}{2} H^2_1 H^2_2 - \mu H_1 H_2. \label{potential}
\eeq
Looking for the minimum of this potential we have the following
constraints,
\beq
&&v_1 (\mu^2_1 + \lambda_1 v^2_1 + \lambda_3 v^2_2) - \mu v_2=0,\nonumber \\
&&v_2 (\mu^2_2 + \lambda_2 v^2_2 + \lambda_3 v^2_1) - \mu v_1=0.
\label{constarint}
\eeq
Taking
\beq \mu^2_2 < 0,\,\,\,\, \mu_1^2 > 0, \,\,\,\, \mu \ll \mu^2_2,
\label{conditions} \eeq
and considering that the second Higgs doublet $H_2$ is in charge
of the quark masses, the value of its vev must be around $v_2 \sim
10^2$~GeV. In this case, we have
\beqq v_1 \sim \frac{\mu v_2}{\mu^2_1 + \lambda_3v^2_2},\,\,\,
v^2_2 \sim -\frac{\mu^2_2}{\lambda_2}. \label{seeaw} \eeqq
Assuming $\mu_1 \sim 10^2$~GeV, and $\mu \sim 10^2$~GeV$^2$, we
find
\beqq v_1 \sim 1\mbox{~GeV}, \label{vacuum} \eeqq
which gives $f_{\mu \mu} = \frac{m_\mu}{v_1} \sim 0.1$, leading
to the expected contribution to $(g-2)_\mu$,
\beq a^\mu \sim 10^{-9}. \label{cvtg} \eeq

A  scheme like this was recently suggested by Ma and Raidal in
two different scenarios. In the first one\cite{ma1,ma2}, the
Higgs doublet $H_1$ carries lepton number and is used to generate
neutrino masses, for which they need a vev, $v_1 \sim 1$~MeV,
which demands a Higgs, $H_1$, with mass around few TeV's. In the
second scenario\cite{ma3}, this Higgs doublet is associated with a
global symmetry $U(1)$ which only permits the doublet to interact
with the light quarks, delegating to $H_1$ the role of generating
their masses. Such a scenario could as well be realized in the
leptonic sector, as we suggested above.

In principle this case also leads to flavor changing processes
whose sources are the interactions: $f_{ij}\bar \nu_{i_L} e_{j_R}
H^+$ for $i\neq j$. However, it is possible to avoid these
processes by assuming lepton number conservation under the
symmetry $U(1)_{L_e,L_\mu, L_\tau}$.  In this case the existing
experimental data are not sufficient to give any constraint over
this scenario. The reason is that a considerable enhancement of
Yukawa couplings only happens for the heaviest leptons, namely,
$f_{\mu \mu}$  and $f_{\tau \tau}$. The enhancement of $f_{ee}$
is not sensible enough to be prompted in an electron collider.
Nevertheless, a scenario like this can be appropriately tested in
a muon collider, where simple processes like $\mu^+ \mu^-
\rightarrow \mu^+ \mu^-\,,\,\, \bar{b}b$ are able to assess such
enhanced Yukawa couplings. An analysis in this direction is done
in Ref.~\cite{mucollider}.

\section{ embedding of both scenarios in a $3-3-1$ gauge theory}
\label{sec3}

Now let us think about a model which naturally accommodates both
scenarios presented above. Such a candidate could be the model
based in the $3-3-1$ symmetry~\cite{331}. In its minimal version
the masses of charged leptons and neutrinos are generated by a
sextet of scalars,
\beq S= \left (
\begin{array}{lcr}
\sigma_1 & \frac{h_2^-}{\sqrt{2}} & \frac{h_1^+}{\sqrt{2}} \\
\frac{h_2^-}{\sqrt{2}} & H_1^{--} & \frac{\sigma_2}{\sqrt{2}} \\
\frac{h_1^+}{\sqrt{2}} & \frac{\sigma_2}{\sqrt{2}} & H^{++}_2
\end{array}
\right ).
\label{sextet}
\eeq

After the breaking of the $3-3-1$ symmetry to the standard $3-2-1$
symmetry, this sextet will decompose under $3-2-1$ in the
following  triplet,  doublet and  singlet of scalars,
respectively~\cite{pires,pleitez}:
\beq
\Delta=
\left (
\begin{array}{lcr}
\sigma_1 & \frac{h_2^-}{\sqrt{2}} \\
\frac{h_2^-}{\sqrt{2}} & H_1^{--}
\end{array}
\right ),\,\,\,\,\,\, \Phi_3=\frac{1}{\sqrt{2}}\left (
\begin{array}{c}
h_1^+ \\
\sigma_2
\end{array}
\right),\,\,\,\,\,\, H^{++}_2 .\label{decomposed} \eeq
Their Yukawa interactions with leptons are
\beq {\cal L}^Y = f_{ii} \bar L^c_{iL} \Delta L_{iL} + f_{ii} \bar
L_{iL}\Phi_3 l_{iR} + f_{ii}\bar l^c_{iR} l_{iR} H_2^{++}.
\label{syukawa} \eeq
Note that the last two interactions in Eq.(\ref{syukawa}) account
for the two scenarios previously discussed.

It was shown in Ref.~\cite{pires} that the vev of the scalars
$\sigma_1$ and $\sigma_2$ are related by
\beqq v_{\sigma_1}=\frac{M v^2_{\sigma_2}}{v^2_\chi}, \label{331}
\eeqq
where $M$ is a free parameter with mass dimension, and $v_\chi$ is
the vev of the triplet $\chi$  which breaks the symmetry $3-3-1$
to $3-2-1$.

Observe that $v_{\sigma_1}$ and $v_{\sigma_2}$ have the same
origin, the sextet. Then we would expect that both take almost the
same value. Nevertheless, one is responsible for the neutrino
masses, and then should be of the order of eV, and the other is
responsible by the charged lepton masses, which are of the order
of GeV. Then it is reasonable to set the value of $v_{\sigma_2}$
around few GeV's. In Eq.(\ref{331}), $v_{\chi}$ is the vev of the
scalar triplet $\chi$, which sets the scale of $3-3-1$ breaking,
around TeV, and $M$ comes from a term in the potential that
violates explicitly the lepton number conservation and should be
small. So, taking the set of values used in Ref.~\cite{pires},
$v_{\sigma_2}=1$~GeV, $v_{\chi}=10$~TeV and $M=0.1$~GeV we obtain
$v_{\sigma_1} =1$~eV, which automatically provides an explanation
for the pointed deviation in $(g-2)_\mu$ and also leads to the
expected scale for neutrino masses (see Ref.~\cite{long} for other
sources contributing to $(g-2)_\mu$ inside $3-3-1$ models).

\section{conclusions}
\label{sec4}

In this work we showed that contributions from scalars with
masses around hundreds of GeV's can contribute significantly to
$(g-2)_\mu$, and have potential capacity of explaining its
theoretical deviation from the observed BNL measurement. We
suggested two scenarios where such scalars would appear
naturally, and a model where they are a constituent part, the
$3-3-1$ gauge model. Besides, both scenarios often appear in
models beyond SM, like neutrino mass models, Left-Right models,
grand unified theories, etc., and so are well motivated.

We should stress though that our analysis is based on one-loop
calculations only. It was shown that in some models containing
scalars (pseudo-scalars), two loop contributions to $(g-2)_\mu$
can be even more significant than the one loop contribution (see
Ref.~\cite{china} for a detailed discussion). This is due to the
Barr-Zee two loop diagram~\cite{bzee}. Essentially what happens
is the following: from the Barr-Zee diagram comes a factor
$m_f/m_\mu$ which can enhance the two loop contributions in
relation to the one loop when a heavy fermion, $f$, is flowing in
the inner loop. The enhancement also depends on other parameters
as $\tan \beta$ and the pseudo-scalar mass, $M_a$, involved in the
loop. According to Ref.~\cite{china} the Barr-Zee type
contribution gives a good fit to the recent 2.6 sigma deviation
when  $\tan \beta \simeq 50$  and $M_a < 40$~GeV. It is not
obvious that such a conclusion would survive for the two Higgs
doublet scenario studied here, since these parameters have values
a little diverse when roughly set according to our scenario,
$\tan \beta \simeq 10^2$ (according to the vev's assumed in
section~\ref{sec2}) and $M_a \simeq 10^2$~GeV. It would be
necessary a thoroughly new calculation to achieve any additional
result besides those already presented in section~\ref{sec2},
which we cannot assure will be relevant. In summary, the simple
scenarios we proposed for heavy scalars and enhanced Yukawa
couplings, can easily accommodate the 2.6 sigma $(g-2)_\mu$
deviation at one loop level.

%%%%%%%%%%%%%%%%%%%%%%%%%%%%%%

{\it Acknowledgements.}  Work supported by Funda\c c\~ao de Amparo \`a
Pesquisa do Estado de S\~ao Paulo
(FAPESP).
%%%%%%%%%%%%%%%%%%%%%%%%%%%%%%

\end{document}